\documentstyle[12pt,axodraw]{article}           

\def\preprint{IFT-07-2001}       
\def\finished{July 2001}
\def\archive {hep-th/0102XXX}           

\def\title{Obtaining a light-like planar gauge.}

\long\def\abstract{
In the usual and current understanding of planar gauge choices for
Abelian and non Abelian gauge fields, the external defining vector
$n_\mu$ can either be space-like ($n^2<0$) or time-like ($n^2>0$) but
not light-like ($n^2=0$). In this work we propose a light-like planar
gauge that consists in defining a modified gauge-fixing term,
$\cal{L}_{GF}$, whose main characteristic is a two-degree violation of
Lorentz covariance arising from the fact that four-dimensional
space-time spanned entirely by null vectors as basis necessitates two
light-like vectors, namely $n_\mu$ and its dual $m_\mu$, with
$n^2=m^2=0,\: n\cdot m\neq 0$, say, e.g. normalized to $n\cdot m=2$.}

\input{epsf}

\begin{document}

{\hfill \archive   \vskip -2pt \hfill\preprint }
\vskip 15mm
\centerline{\huge\bf Obtaining a light-like}
\centerline{\huge\bf planar gauge.}
\vskip 8mm
\begin{center} \vskip 10mm
Alfredo T. Suzuki{\begingroup\def\thefootnote{a}
                                \footnote{e-mail: suzuki@ift.unesp.br}
                                \addtocounter{footnote}{-1} \endgroup}
and \  \
Ricardo Bent\'{\i}n{\begingroup\def\thefootnote{z}
                                \footnote{e-mail: rbentin@ift.unesp.br}
                                \addtocounter{footnote}{-1} \endgroup}
\\[3mm] Instituto de F\'{\i}sica Te\'orica - UNESP.\\
        Rua Pamplona 145, CEP 01405-900, SP, Brasil.
        
\vfill                  {\bf ABSTRACT } \end{center}    \abstract

\vfill \noindent \preprint\\[5pt] \finished \vspace*{9mm}
\thispagestyle{empty} \newpage
\pagestyle{plain} 

\newpage
\setcounter{page}{1}

\baselineskip1cm

\section{\large Introduction.}
Gauge fields regardless of their Abelian or non Abelian character are
by definition fields that display some sort of gauge
invariance. Fields which are connected by this invariance must be
accounted only once in the process of quantization, otherwise all sort
of pathologies arise. The standard and pragmatic way in which one
proceeds in dealing with this issue is the introduction of a
gauge-breaking term in the Lagrangian density which fixes the gauge
choice. This choice is obviously a matter of taste and convenience of
one's preference, although practice has taught us that some choices
may bring more technically demanding difficulties than others.

In this respect, algebraic non-covariant gauges have posed many
challenges concerning not only the question of their usefulness but
also what concerns the inherent technical difficulties and subleties
associated with them. For example, an open problem in the
non-homogeneous axial gauge, also known as {\em planar} gauge
\cite{dok,kum}, has to do with the fact that {\em ``it is not 
possible''}
to define it in the light-like case \cite{bass}. Let us briefly review
the reason why this is considered so.

The non-homogeneous axial case is defined by the condition:
$$
  n^\mu A^a_\mu\equiv n\cdot A^a=\Phi^a
$$
where $n_\mu$ is the arbitrary external and constant vector, $A^a_\mu$
are the gauge fields (for definiteness we deal with general
non-Abelian fields) and $\Phi^a$ are a set of scalar fields belonging
to the adjoint representation of $SU(N)$ Lie algebra \cite{bass}. Then
the gauge fixing Lagrangian density term is chosen to be:

\begin{equation}
  \label{usual01}
  {\cal L}_{GF}=\frac{1}{2n^2} \partial_\mu\Phi^a\partial^\mu\Phi^a
  +\lambda^a(\Phi^a-n\cdot A^a).
\end{equation}
where $\lambda^a$ are the Lagrange multipliers. It is easy to see
that these last equations are in fact equal to:
\begin{equation}
  \label{usual02} {\cal L}_{GF}=\frac{1}{2n^2} \partial_\mu(n\cdot
  A^a)\partial^\mu(n\cdot A^a).
\end{equation}
Depending on the type of external vector $n_\mu$ that we consider the
following cases are defined:
\begin{itemize}
\item $n^2<0$: space-like planar gauge,
\item $n^2>0$: time-like planar gauge.
\end{itemize}

However, the light-like planar gauge is not defined, since in this
case we meet a singularity at $n^2=0$. 

But in the next section we will see that there is a way out to this
inconvenience.
\section{\large Our proposal.}
We begin this section by reviewing some important concepts related to
null vectors as basis for four dimensional space-time. The
``light-likeness'' condition $n^2=0$ does not uniquely define the
necessary external vector $n_\mu$ to implement the gauge
condition. The reason for this is most easily seen considering a
particular case where $n_\mu=(n_0,\;0,\;0,\;n_3)$, in which case the
condition $n^2=0$ gives as solutions either $n_0=+n_3$, or $n_0=-n_3$
with $n_0>0$. Therefore those components of the light-like vector are
not linearly independent; hence the two possibilies are
$n_\mu=(n_0,\;0,\;0,\;+n_3)$ and $m_\mu=(n_0,\;0,\;0,\;-n_3)$. These
peculiar light-cone vector properties have been demonstrated by
Leibbrandt \cite{Leib} to be connected to the Newman-Penrose
\cite{Penr} tetrad formalism in the context of gravitation and
cosmology, where a four-dimensional basis is spanned entirely by null
vectors. In his work, Leibbrandt demonstrated that the two-dimensional
vector sub-space $(n_0,\;n_3)$ {\em cannot} be spanned solely by the
vector $n_\mu$, because this vector possesses linearly dependent
components.

Therefore, for our purpose and without loss of generality, we
introduce two light-like vectors, namely $n_\mu$ and $m_\mu$, so that
(note that here we choose a normalization factor 2 for convenience)

\begin{eqnarray*}
  n^2=m^2&=&0,\\
  n\cdot m&=&2.
\end{eqnarray*}

Our proposal is to consider the following gauge fixing term
\begin{eqnarray}
  \label{new01}
 {\cal L}_{GF}=\frac{1}{4n\cdot 
m}\partial_\mu\Phi^{2a}\Sigma\partial^\mu\Phi^{2a}
  +\lambda^{2a}(\Phi^{2a}-{\cal N}\cdot A^{2a}).
\end{eqnarray}
where
\begin{eqnarray*}
  \Sigma&=&       \left(
                        \begin{array}{cc}
                                   0 & 1 \\
                                   1 & 0 \\
                        \end{array}
                  \right)
\end{eqnarray*}
and
\begin{eqnarray*}
  {\cal N}_\mu&=&     \left(
                        \begin{array}{cc}
                                   1\otimes n_\mu & 0 \\
                                   0 & 1\otimes m_\mu \\
                        \end{array}
                  \right)
\end{eqnarray*}

Also we extend the dimension $N$ of the Lie algebra to be $2N$ 
or it is extended to the complex case, so that we can split the
$\Phi^{2a}$ ($2a=1\cdots 2N$) fields as:

\begin{eqnarray*}
  \Phi^{2a}&=&     \left(
                        \begin{array}{c}
                                   \phi^a \\
                                   \varphi^b\\
                        \end{array}
                  \right)
\end{eqnarray*}
where $a=1\cdots N$ and $b=N+1\cdots 2N$ in the Lie algebra space.  
It is not too hard to
check that the gauge fixing term described in equation (\ref{new01})
will take the form

\begin{equation}
  \label{new02} {\cal L}_{GF}=\frac{1}{2n\cdot m} \partial_\mu(n\cdot
  A^a)\partial^\mu(m\cdot A^a),
\end{equation}
just working it through as it was done when going from (\ref{usual01}) 
to (\ref{usual02}) (where we have made the ansatz $A^a=A^b$).

This last equation is in fact free from the singularity in the
light-like case since now we have $n\cdot m$ in the denominator and
this quantity differs from zero. So it is possible to define
(\ref{new01}) to be the gauge fixing Lagrangian density term for the
planar gauge in the light-like case.\\

Note that there is a kind of discrete simmetry in (\ref{new02}) when
we exchange $n\leftrightarrow m$. Making such an exchange
(\ref{new01}) results in the ``breaking'' of the matrix ${\cal N}$, so
(\ref{new01}) is not invariant under this kind of simmetry, but if
we use this broken gauge fixing Lagrangian, and then procede to
eliminate the $\Phi^a$ fields as it was done when going from
(\ref{new01}) to (\ref{new02}) then we will discover that our answer
coincides with (\ref{new02}).  

With this new gauge fixing term we can find the boson propagator,
\begin{eqnarray}
\nonumber
 {\cal  G}_{\mu\nu}(q)&=&-{\frac{i}{2(2{\pi})^4q^2}}
        {\biggl[}2g_{\mu \nu}
        -q^-{\frac{(q_{\mu}n_{\nu}+q_{\nu}n_{\mu})}{q^+q^-}}
        -q^+{\frac{(q_{\mu}m_{\nu}+q_{\nu}m_{\mu})}{q^+q^-}}
        +{\frac{q_{\mu}q_{\nu}}{q^+ q^-}}\\
        &&+{\frac{(q^-)^2}{q^+q^-}}n_{\mu}n_{\nu}
        +{\frac{(q^+)^2}{q^+q^-}}m_{\mu}m_{\nu}
        -q^+q^-{\frac{(n_{\mu}m_{\nu}+m_{\mu}n_{\nu})}{q^+q^-}}
        {\biggr]} \label{gnm}\\
\nonumber
        &\equiv&-{\frac{i}{2(2{\pi})^4q^2}}\left[ G_{\mu\nu}^{\sc
        cov}(q)+G_{\mu\nu}^{\sc planar}(q,n,m)\right],
\end{eqnarray}

Observe that the vector propagator now is much more complex than
the usual one which is obtained using only one external vector $n_\mu$
to fix the gauge choice (one-degree violation of Lorentz
covariance). On the other hand, it is clear from this new propagator
that the light-cone variables $q^+$ and $q^-$ do form a bilinear term
$q^+q^-$ that cannot be separated without loosing important physical
information in the process \cite{cov,ndim}. In fact, separating them in 
the
denominators means violating causality for the field operators. So,
although one may be tempted to simplify the above expression into

\begin{eqnarray*}
  G_{\mu\nu}^{\sc planar}(q,n,m)=
        -{\frac{(q_{\mu}n_{\nu}+q_{\nu}n_{\mu})}{q^+}}
        -{\frac{(q_{\mu}m_{\nu}+q_{\nu}m_{\mu})}{q^-}}
        +{\frac{q_{\mu}q_{\nu}}{q^+ q^-}}\\
        +{\frac{q^-}{q^+}}n_{\mu}n_{\nu}
        +{\frac{q^+}{q^-}}m_{\mu}m_{\nu}
        -n_{\mu}m_{\nu}-m_{\mu}n_{\nu},
       \end{eqnarray*}

\noindent this simplification may pose unsuspected mixing of positive- 
and 
negative-frequency modes for the quantum fields.

\section{\large Conclusions.}

In this work we have considered the algebraic noncovariant gauge
choice of the inhomogeneouos type --- known as planar gauge ---
introducing a two-degree violation of Lorentz covariance on the gauge
breaking term in the Lagrangean density, supported by the fact that
four-dimensional space-time spanned solely by null vectors as basis
cannot be accomplished without the concurrence of both pairs of dual
light-like vectors $n_\mu$ and $m_\mu$. Doing this, we were able to
define the planar gauge in the light-like case, however, the new boson
vector propagator becomes almost prohibitively complicated for
calculating Feynman integrals in the diagrams of quantum corrections
to any physical processes involving them. 

\vspace{.5cm}

{\bf Acknowledgments:} R.B. wishes to thank {\sc fapesp} for 
financial support.

\end{document}